\documentclass[aps,prb,twocolumn,superscriptaddress,showpacs,10pt]{revtex4-1}
\usepackage{amsmath}
\usepackage{amssymb}
\usepackage{bm}
\usepackage[utf8]{inputenc}
\usepackage{mathptmx}
\usepackage{graphicx}
\usepackage{multirow}
\usepackage{subfigure}
\usepackage[bookmarks=true,colorlinks=true,linkcolor=blue,citecolor=blue,urlcolor=blue]{hyperref}

\newcommand{\vect}[1]{\bm{#1}}
\newcommand{\argu}[1]{\!\left( {#1} \right)}
\newcommand{\set}[1]{\left\lbrace {#1} \right\rbrace}
\newcommand{\avg}[1]{\left\langle {#1} \right\rangle}
\newcommand{\bra}[1]{\left\langle {#1} \right\rvert}
\newcommand{\ket}[1]{\left\lvert {#1} \right\rangle}
\DeclareMathOperator{\im}{Im}
\DeclareMathOperator{\Tr}{Tr}
\DeclareMathOperator{\e}{e}

\begin{document}

\title{Metallic magnetism at finite temperatures studied by relativistic disordered moment description:\\Theory and applications}

\author{A. De\'ak}
\email{adeak@phy.bme.hu}
\affiliation{Department of Theoretical Physics, Budapest University of Technology and Economics, Budafoki \'ut 8., HU-1111 Budapest, Hungary}

\author{E. Simon}
\affiliation{Department of Theoretical Physics, Budapest University of Technology and Economics, Budafoki \'ut 8., HU-1111 Budapest, Hungary}

\author{L. Balogh}
\affiliation{Department of Theoretical Physics, Budapest University of Technology and Economics, Budafoki \'ut 8., HU-1111 Budapest, Hungary}

\author{L. Szunyogh}
\affiliation{Department of Theoretical Physics, Budapest University of Technology and Economics, Budafoki \'ut 8., HU-1111 Budapest, Hungary}
\affiliation{Condensed Matter Research Group of the Hungarian Academy of Sciences, Budapest University of Technology and Economics, Budafoki \'ut 8., HU-1111 Budapest, Hungary}

\author{M. dos Santos Dias}
\affiliation{Peter Grünberg Institut and Institute for Advanced Simulation, Forschungszentrum Jülich and JARA, D-52425 Jülich, Germany}

\author{J. B. Staunton}
\affiliation{Department of Physics, University of Warwick, Coventry CV4 7AL, United Kingdom}

\date{\today}

\begin{abstract}
We develop a self-consistent relativistic disordered local moment (RDLM) scheme aimed at describing finite temperature magnetism of itinerant metals from first principles. Our implementation  in terms of the Korringa--Kohn--Rostoker multiple scattering theory and the coherent potential approximation allows to relate the orientational distribution of the spins to the electronic structure, thus a self-consistent treatment of the distribution is possible. We present applications for bulk bcc Fe, L1$_0$-FePt and FeRh ordered in the CsCl structure. The calculations for Fe show significant variation of the local moments with temperature, whereas according to the mean field treatment of the spin fluctuations the Curie temperature is overestimated. The magnetic anisotropy of FePt alloys is found to depend strongly on intermixing between nominally Fe and Pt layers, and it shows a power-law behavior as a function of magnetization for a broad range of chemical disorder. In case of FeRh we construct a  lattice constant vs.\ temperature phase diagram and determine the phaseline of metamagnetic transitions based on  self-consistent RDLM free energy curves.
\end{abstract}

\pacs{71.15.Mb, 71.15.Rf, 75.30.Gw, 75.30.Kz}

\maketitle

\section{Introduction}
\label{intro}
Magnetic metals are typified by a strong interrelation between the electronic states and magnetic ordering.
Effects of temperature-induced magnetic fluctuations are of special interest, since respective changes 
in the electronic structure sensitively contribute to the temperature dependence of important physical properties 
like the geometric structure, the electric, optical or spin transport of the system. 
In particular, if a metallic magnet passes through a first order magnetic transition the changes to the electronic structure 
can be significant and the mentioned magnetic effects can be dramatic.  A prototypical example is evidenced 
by the metamagnetic phase transition of FeRh in connection to a large magnetocaloric effect. \cite{fallot,annaorazov,driel,kobayashi,maat,stamm,suzuki,vries}  
Another intriguing phenomenon is the non-trivial temperature dependence of the magnetic anisotropy energy (MAE) of ordered and disordered FePt alloys. \cite{okamoto-2002,thiele-2002,staunton-2004,mryasov-2005a}

First principles studies of magnetism at finite temperatures go back to the work of Mermin \cite{mermin} who extended density functional theory (DFT) to include the statistical distribution of effective non-interacting electrons. This theory failed, however, to reproduce the Curie temperature of elementary ferromagnets, since only the high energy Stoner excitations were taken into account. A series of subsequent theoretical works\cite{korenmann-1977,hubbard-1979,hasegawa-1979, edwards-1982,moriya-1985,gyorffy-1985} 
reached the consensus that thermal properties of metallic magnets with strong local moments are governed rather by orientational 
fluctuations of the local magnetization at the energy scale  comparable to the Curie temperature.  
Ab initio theories of spin fluctuations are based on the notion of adiabatic spin-dynamics without relying on a model spin Hamiltonian, for an illuminating overview see Ref.\ \onlinecite{kubler-2000}.  

Spin-density functional theory (SDFT) has been merged with the disordered local moment (DLM) scheme by Gy\"orffy \emph{et al.}\cite{gyorffy-1985}\ treating spin fluctuations within a mean field approximation. It has been shown that DLM as implemented with the Korringa--Kohn--Rostoker (KKR) multiple scattering method and the coherent potential approximation (CPA) provides a feasible tool for calculating the electronic structure in the presence of fluctuating local spin moments. This theory was first applied to the paramagnetic state of ferromagnetic metals in a non-relativistic setting, where owing to the rotational symmetry of the paramagnetic state the calculations can be mapped to those of an Ising type system of up and down moments of equal distribution.\cite{gyorffy-1985,staunton-1986} The relativistic extension of DLM (RDLM) was then introduced to calculate the temperature dependence of the magnetic anisotropy of bulk and thin film systems.\cite{staunton-2004,staunton-2006,buruzs-2007}   
The RDLM scheme for temperatures below the paramagnetic transition temperature has recently been employed to study metamagnetism in antiferromagnetic alloys \cite{staunton-2013} and the magnetocaloric effect in compositionally disordered FeRh alloys. \cite{staunton-2014} 

In almost all previous applications of the RDLM scheme the effect of the spin-disorder on the effective potentials, in particular to the local spin-polarization (exchange splitting) was, however, neglected, and the effective potentials and fields obtained in either the ferromagnetic ($T=0$~K) or in the paramagnetic (DLM) state were used.  This approach relies on the original notion of ``good moments'' characteristic to ferromagnets like Fe or Co, but certainly doesn't apply to Ni. In particular, in magnetic alloys, like FePt, FeRh and many others, containing atoms with induced moments generated by the strong spin moments, the interplay between the local exchange splitting and the transversal spin fluctuations is essential. Interpolation between paramagnetic and ferromagnetic self-consistent potentials was used for the Co/Cu(100) thin film system to demonstrate the sensitivity of the magnetic anisotropy energy to the choice of the potentials.\cite{buruzs-2007}

In this paper we extend calculations within the RDLM scheme by updating the Kohn--Sham potentials and exchange fields self-consistently. This development allows to calculate the local exchange splitting of the fluctuating spins as a function of temperature or average magnetization. For completeness, in Section \ref{theory} we summarize the main features of the RDLM theory together with the above extension, pointing out the approximations we used in the actual implementation. Special attention is devoted to the calculation of the free energy including the electronic contribution.  In Section \ref{results} we first test the method on bulk Fe and demonstrate the dependence of the local magnetic moments on temperature-induced spin fluctuations. Then we perform calculations on ordered and disordered bulk FePt alloys focusing on the temperature dependence of the magnetic anisotropy energy. Finally we study the antiferromagnetic (AFM) and ferromagnetic (FM) phases of ordered FeRh alloys. We derive an \emph{ab initio} phase diagram in terms of the lattice constant and the temperature by finding evidence of metamagnetic phase transitions in reliable agreement with experiments and previous calculations.  In Section \ref{conclusions} we summarize and outline further extensions and applications of the theory. 
 
\section{RDLM theory}
\label{theory}
DLM theory and its implementation within the Korringa--Kohn--Rostoker multiple scattering theory was given by Gy\"orffy \emph{et al.},\cite{gyorffy-1985} with a relativistic generalization by Staunton \emph{et al.}\cite{staunton-2004,staunton-2006} The DLM scheme describes a magnetic system as a set of fluctuating local moments within the adiabatic approximation, according to which slow spin degrees of freedom are decoupled from the fast (electronic) degrees of freedom. In this approximation it is meaningful to assume a set of unit vectors $\left\lbrace \vect e\right\rbrace=\left\lbrace \vect e_1, \vect e_2, \dots \right\rbrace$ describing the spin configuration of the fluctuating system. RDLM theory describes the fluctuations of the finite-temperature system in terms of single-site probabilities, inherently providing a local mean field description of spin disorder. Besides the spin disorder, chemical disorder can be treated on an equal footing in terms of the coherent potential approximation. 

Within the DLM theory the statistical probability of the disordered spin system is approximated by independent single-site concentrations and orientational probabilities,
\begin{align}
\mathbb P\argu{\set{\xi},\set{\vect e}}=\prod_i \sum\limits_{\alpha} \xi_{i\alpha} c_{i\alpha} P_{i\alpha}\argu{\vect e_{i\alpha}}
\label{eq:P}
\end{align}
where $\set{\xi}$ and $\set{\vect e}$ describe a specific chemical and orientational configuration, respectively. Here $c_{i\alpha}$ is the probability of finding a chemical component of type $\alpha$ at site $i$, $\vect e_{i\alpha}$ is the spin direction of component $\alpha$ at the same site, and $\xi_{i\alpha}$ are binary random variables for chemical species $\alpha$ at site $i$, i.e. $\xi_{i\alpha}=1$ if site $i$ is occupied by species $\alpha$, otherwise it is zero. The single-site orientational probability densities are sought for as canonical distributions at temperature $T$,
\begin{align}
P_{i\alpha}\argu{\vect e_{i\alpha}}=\frac{1}{Z}\e^{-\beta h_{i\alpha}\argu{\vect e_{i\alpha}}},\label{eq:meanfield_P}
\end{align}
where $Z$ is the canonical partition function and $\beta^{-1}=k_B T$. The exponent $h_{i\alpha}\argu{\vect e_{i\alpha}}$ is chosen to give the best approximation of the disordered system. This should be determined by the Feynman--Peierls--Bogoliubov inequality,
which relates the free energy ($F$) corresponding to the Hamiltonian of interest ($H$) to the free energy ($F_0$) of an approximating trial Hamiltonian ($H_0$):
\begin{align}
F\leqslant F_0 + \avg{H-H_0},
\end{align}
where the average has to be taken with the canonical distribution corresponding to $H_0$. 
For a mean field (i.e., single-site) trial Hamiltonian,
\begin{align}
H_0(\set{\xi},\set{\vect e}) = \sum_{i,\alpha} \xi_{i\alpha} h_{i\alpha}\argu{\vect e_{i\alpha}} \, ,
\end{align}
the optimal parameters are given by the conditional average\cite{gyorffy-1985}
\begin{align}
h_{i\alpha}\argu{\vect e_{i\alpha}}=\avg{H\argu{\set{\xi},\set{\vect e};\vect B_\text{ext}}}_{\vect e_{i\alpha}},\label{eq:variational_h}
\end{align}
for which the chemical species and its spin is kept fixed at site $i$ during averaging. 
In general the Hamiltonian entering Eq.~\eqref{eq:variational_h} may contain an external field $\vect B_\text{ext}$ to allow for the computation of response functions. 

For a given chemical and orientational configuration the electronic charge and magnetization densities are determined from a self-consistent field (scf) KKR calculation. In principle one has to perform a constrained local moment density-functional theory (CLM-DFT) calculation\cite{stocks-1998,ujfalussy-1999}
 with every possible set of $\set{\xi}$ and $\set{\vect e}$. 
Within the KKR Green's function method this provides us with the charge density\cite{zabloudil-2005}
\begin{align}
\rho\argu{\vect r;\set{\xi},\set{\vect e}}&=\mathfrak I \Tr \bra{\vect r} G\argu{\varepsilon; \set{\xi},\set{\vect e}}\ket{\vect r}
\end{align}
for each configuration, with the $G\argu{\varepsilon;\set{\xi},\set{\vect e}}$ resolvent of the system for energy $\varepsilon$. Here we introduced a simplified notation
\begin{align}
\mathfrak I g=-\frac{1}{\pi}\im \int f\argu{\varepsilon;\mu} g\argu{\varepsilon} \mathrm d\varepsilon
\end{align}
for the ubiquitous energy integrals containing the $f\argu{\varepsilon;\mu}$ Fermi-function.

Within the RDLM scheme the conditional average of these charge densities 
\begin{align}
\rho_{i\alpha}\argu{\vect r_i;\vect e_{i\alpha}}&=\mathfrak I \Tr \bra{\vect r_i} \avg{G\argu{\varepsilon; \set{\xi},\set{\vect e}}}_{\vect e_{i\alpha}} \ket{\vect r_i}
\end{align}
 is used at site $i$ for chemical species $\alpha$. Similarly, the conditional average of the longitudinal component of the magnetization density is given by
\begin{align}
m_{i\alpha}\argu{\vect r_i;\vect e_{i\alpha}}&=\mathfrak I \Tr \bra{\vect r_i} \vect e_{i\alpha}{\cdot} \beta \vect\Sigma \avg{ G\argu{\varepsilon; \set{\xi},\set{\vect e}}}_{\vect e_{i\alpha}} \ket{\vect r_i} ,
\end{align}
with the usual 4$\times$4 matrices $\beta$ and $\vect\Sigma$, within a relativistic formalism.\cite{eschrig-1996} 
Using these average densities one obtains the chemical species and spin direction dependent effective potentials and exchange fields,
\begin{align}
V_{i\alpha}\argu{\vect r_i;\vect e_{i\alpha}}&=V\left[\rho_{i\alpha}\argu{\vect r_i;\vect e_{i\alpha}},m_{i\alpha}\argu{\vect r_i;\vect e_{i\alpha}}\right],\\
B_{i\alpha}\argu{\vect r_i;\vect e_{i\alpha}}&=B\left[\rho_{i\alpha}\argu{\vect r_i;\vect e_{i\alpha}},m_{i\alpha}\argu{\vect r_i;\vect e_{i\alpha}}\right].
\end{align}
The solution of the Dirac equation with these potentials for a given energy $\varepsilon$ determines the configuration-dependent single-site $t$-matrices $ \underline t_{i\alpha}\argu{\varepsilon;\vect e_{i\alpha}}$ which are the basic quantities in KKR describing the single-site scattering problem (underlines denote matrices in the $\left(\kappa,\mu\right)$ angular momentum representation). The energy arguments of the appearing matrices will be omitted in the following.

The local CPA is employed to describe the disordered system, in accordance with the mean field nature of the probability density. The strategy of the local CPA is to substitute the disordered system with an effective (coherent) medium, characterized by the coherent $t$-matrices, $ \underline t_{c,i}$, which are independent from the orientation of local moments and the chemical configuration, such that the scattering of an electron in the effective medium should resemble the average scattering in the disordered physical system. As the central quantity of the KKR Green's function formalism, the matrix of the scattering path operator of the effective medium is defined as\cite{zabloudil-2005}
\begin{align}
\underline{\underline \tau}_c=\left(\underline{\underline t}_c^{-1}-\underline{\underline G}_0\right)^{-1},
\end{align}
where double underlines denote matrices in site-angular momentum space, $\underline{\underline G}_0$ is the matrix of structure constants, and $\underline{\underline t}_c$ is site diagonal.
The single-site CPA condition can then be formulated as
\begin{align}
\underline\tau_{c,ii}&=\sum\limits_\alpha c_{i\alpha}\int \bigl\langle \underline\tau_{i\alpha,i\alpha} \argu{\set{\xi},\set{\vect e}}\bigr\rangle_{\vect e_{i\alpha}} P_{i\alpha}\argu{\vect e_{i\alpha}}\,\mathrm d^2 e_{i\alpha},
\end{align}
or by introducing the excess scattering matrices
\begin{align}
\underline X_{i\alpha}\!\left(\vect e_{i\alpha}\right)=\left[\left(\underline t_{c,i}^{-1}-\underline t_{i\alpha}^{-1}\!\left(\vect e_{i\alpha}\right)\right)^{-1} - \underline \tau_{c,ii}\right]^{-1},
\end{align}
as
\begin{align}
\sum\limits_\alpha c_{i\alpha}\int P_{i\alpha}\argu{\vect e_{i\alpha}} \underline X_{i\alpha}\!\left(\vect e_{i\alpha}\right)\, \mathrm d^2 e_{i\alpha} =\underline 0 \: .
\end{align}
The CPA condition has to be solved self-consistently along with the probability densities describing spin disorder. The single-site Hamiltonian $h_{i\alpha}\argu{\vect e_{i\alpha}}$ can be expanded as
\begin{align}
h_{i\alpha}\argu{\vect e_{i\alpha}}=\sum\limits_L h_{i\alpha}^L Y_L\argu{\vect e_{i\alpha}},\label{eq:h_expansion}
\end{align}
where the $Y_L$ stand for real spherical harmonics with composite quantum number $L=\left(\ell,m\right)$. The expansion coefficients $h_{i\alpha}^L$ have to be chosen according to Eq.~\eqref{eq:variational_h}, where the role of the Hamiltonian of the disordered system should be played by the grand potential of the system,\cite{gyorffy-1985}
\begin{align}
\Omega\argu{\set{\xi},\set{\vect e}} = E_\text{tot}\argu{\set{\xi},\set{\vect e}} - \epsilon_F N\argu{\set{\xi},\set{\vect e}}  \; ,
\end{align}
where $E_\text{tot}\argu{\set{\xi},\set{\vect e}}$ and $N\argu{\set{\xi},\set{\vect e}}$ are the total energy and
the integrated density of states (DOS) for a given configuration of the system, respectively, while $ \epsilon_F$ is the Fermi energy.
The total energy of the system is given within the SDFT as\cite{zabloudil-2005}
\begin{align}
E_\text{tot}\argu{\set{\xi},\set{\vect e}}=E_\text{kin}+E_\text{H}+E_\text{ext}+E_\text{xc},\label{eq:etot}
\end{align}
where $E_\text{kin}$ is the kinetic energy, $E_\text{H}$ is the Hartree energy,  $E_\text{xc}$ is the exchange-correlation energy and $E_\text{ext}$ is the energy of external potentials and magnetic fields (including the contributions from the nuclear potential). By using the Kohn--Sham equations, the kinetic energy $E_\text{kin}$ can be decomposed as
\begin{align}
E_\text{kin}=E_\text{s}&-\sum\limits_{i\alpha} \xi_{i\alpha} \int\limits V_{i\alpha}\argu{\vect r_i;\vect e_{i\alpha}} \rho_{i\alpha}\argu{\vect r_i;\vect e_{i\alpha}}\mathrm d^3 r_i\notag\\
&-\sum\limits_{i\alpha} \xi_{i\alpha} \int\limits \vect B_{i\alpha}\argu{\vect r_i;\vect e_{i\alpha}} \vect m_{i\alpha}\argu{\vect r_i;\vect e_{i\alpha}}\mathrm d^3 r_i,
\end{align}
where $E_\text{s}$ is the single-particle energy:
\begin{align}
E_\text{s}=E_\text{core}\argu{\set{\xi},\set{\vect e}}+E_\text{band}\argu{\set{\xi},\set{\vect e}}.
\end{align}
Here $E_\text{core}\argu{\set{\xi},\set{\vect e}}$ stands for the sum of the energies of the core eigenstates, and the band energy reads as
\begin{align}
E_\text{band}\argu{\set{\xi},\set{\vect e}}=\int\limits f\argu{\varepsilon;\mu} \varepsilon \, n\argu{\varepsilon;\set{\xi},\set{\vect e}} \mathrm d\varepsilon,
\end{align}
with the density of states (DOS)
\begin{align}
n\argu{\varepsilon;\set{\xi}, \set{\vect e}}=\sum_{i,\alpha} \xi_{i\alpha} n_{i\alpha}\argu{ \varepsilon; \vect e_{i\alpha}}.
\end{align} 
Note that every term in Eq.~\eqref{eq:etot} depends on the orientation of the local moments through the densities.
In addition, the single-particle energy, $E_\text{s}$, implicitly depends on $\set{\vect e}$ through the DOS.

Eq.~\eqref{eq:h_expansion} together with Eq.~\eqref{eq:variational_h} prescribes the single-site expansion coefficients as
\begin{align}
h_{i\alpha}^L=\int Y_{L}\argu{\vect e_{i\alpha}} \avg{\Omega\argu{\set{\xi},\set{\vect e}}}_{\vect e_{i\alpha}} \mathrm d^2 e_{i\alpha}.\label{eq:g_i_general}
\end{align}

By finding the relationship between the orientational probability and the electronic structure a self-consistent treatment of spin disorder is possible. A self-consistent field calculation consists of starting from a set of initial probabilities, potentials and exchange fields, performing the CPA to obtain the $t$-matrices and scattering path operator of the coherent medium, calculating the new expansion coefficients using Eq.~\eqref{eq:g_i_general}, then starting a new iteration with the resulting probability densities and potentials. Once convergence of the densities and probabilities is achieved, the required physical quantities can be calculated.

It should be mentioned that in case of induced moments the adiabatic approximation might not be valid at all. For this reason, in our implementation only the good moment constituents are described according to DLM, 
whilst induced moments are treated within usual SDFT. Though the orientation of the induced moments can be
determined self-consistently, for ferromagnetic systems we kept it parallel to the average magnetization of 
the good moments. For antiferromagnetic alloys like FeRh the induced moment of Rh converged to zero.

Since for good moments the magnitude of local moments is usually considered independent from their orientation,
as a further approximation we neglected the direction dependence of the densities 
(and hence of the effective potential and exchange field).  We note that this is just a reasonable computational simplification, but not a methodological necessity as the self-consistent procedure could be performed as described above. The resulting direction averaged densities,
\begin{align}
\rho_{i\alpha}\argu{\vect r_i}&=\mathfrak I \int P_{i\alpha}\argu{\vect e_{i\alpha}}\notag\\
&\times \Tr \bra{\vect r_i} \avg{G\argu{\varepsilon;\set{\xi},\set{\vect e}}}_{\vect e_{i\alpha}} \ket{\vect r_i}\mathrm d^2e_{i\alpha},\label{eq:avgrho}\\
m_{i\alpha}\argu{\vect r_i}&=\mathfrak I\int P_{i\alpha}\argu{\vect e_{i\alpha}} \notag \\
& \times \Tr \bra{\vect r_i} \beta \vect e_{i\alpha}{\cdot}\vect\Sigma \avg{ G\argu{\varepsilon;\set{\xi},\set{\vect e}}}_{\vect e_{i\alpha}} \ket{\vect r_i}\mathrm d^2 e_{i\alpha}\label{eq:avgmag}
\end{align}
can be used in a conventional SDFT calculation to obtain $V_{i\alpha}$ and $B_{i\alpha}$ which now only depend on the chemical species (in effect an averaged Green's function provides the densities). The component- and site-resolved average magnetization,  
\begin{align}
\vect M_{i\alpha}\argu{T}&=\mathfrak I\iint P_{i\alpha}\argu{\vect e_{i\alpha}} \notag \\
& \!\!\!\!\times\! \Tr \bra{\vect r_i} \beta \vect\Sigma \avg{ G\argu{\varepsilon;\set{\xi},\set{\vect e}}}_{\vect e_{i\alpha}} \ket{\vect r_i}\mathrm d^3 r_i \,\mathrm d^2 e_{i\alpha} ,
\end{align}
is zero in the PM phase due to symmetry, whereas its magnitude approaches the size of the local spin moment as the temperature tends to zero. The total magnetization of the system is then given by
\begin{align}
\vect M\argu{T}=\sum\limits_{i,\alpha} c_{i\alpha} \vect M_{i\alpha}\argu{T}.\label{eq:total_mag}
\end{align}

Neglecting the orientational dependence of the densities and effective potentials implies that the direction dependence of the grand potential entering the expansion coefficients, Eq.~\eqref{eq:g_i_general}, comes from the band energy contribution to the single-particle energy only. The band energy part of the grand potential is given by
\begin{align}
\Omega\argu{\set{\xi},\set{\vect e}}\approx -\int f\argu{\varepsilon;\mu} N\argu{\varepsilon;\set{\xi},\set{\vect e}} \mathrm d\varepsilon
\end{align}
using the integrated DOS $N\argu{\varepsilon;\set{\xi},\set{\vect e}}$, and the grand potential of the disordered system can be expressed by making use of the Lloyd formula. Straightforward calculation leads to the expression\cite{szunyogh-2011}
\begin{align}
h_{i\alpha}^L= \mathfrak I \int Y_L\argu{\vect e_{i\alpha}}\ln\det \underline D_{i\alpha}\argu{\vect e_{i\alpha}}\mathrm d^2 e_{i\alpha}\label{eq:g_i}
\end{align}
with the impurity matrix
\begin{align}
\underline D_{i\alpha}\argu{\vect e_{i\alpha}}=\underline I +\underline X_{i\alpha}\argu{\vect e_{i\alpha}} \underline \tau_{c,ii}.
\end{align}

For finite temperatures the relevant thermodynamic potential is the free energy that is defined within the RDLM  scheme as
\begin{align}
F\argu{T}=\left\langle E_\text{tot} \argu{\set{\xi},\set{\vect e},T}\right\rangle-TS_\text{c} -TS_\text{el},
\end{align}
where $\left\langle E_\text{tot} \argu{\set{\xi},\set{\vect e},T}\right\rangle$ is the statistically averaged DFT total energy, $S_\text{c}$ stands for the configurational entropy of the system (both spin and chemical),
\begin{align}
S_\text{c}=-k_B \left\langle \ln \mathbb P\!\left(\set{\xi},\set{\vect e}\right) \right\rangle, \label{eq:Sc}
\end{align}
which can be calculated from Eqs.~\eqref{eq:P} and \eqref{eq:meanfield_P}, and $S_\text{el}$ denotes the electronic entropy. Since the temperature ranges associated with magnetic ordering (for instance, Curie- or Néel-temperatures) are much smaller than the temperature scale of electronic degrees of freedom (i.e., the Fermi temperature), contributions arising from finite electronic temperature can be treated in terms of a Sommerfeld expansion. In our simulations the electronic structure is assumed to be in the ground state (i.e., the Fermi function in the energy integrals is substituted with a step function), correspondingly the free energy we have to use is given by
\begin{align}
F\argu{T}\approx \left\langle E\argu{\set{\xi},\set{\vect e},0}\right\rangle-TS_\text{c} +\Delta F_\text{el}\argu{T}, \label{eq:F}
\end{align}
where $\left\langle E\argu{\set{\xi},\set{\vect e},0}\right\rangle$ denotes the averaged total energy with zero electronic temperature, and
\begin{align}
\Delta F_\text{el}\argu{T}=-\frac{\pi^2}{6} \left(k_B T\right)^2 n\argu{\varepsilon_F} \label{eq:F_el}
\end{align}
is the excess free energy contribution of the electrons at temperature $T$  with the averaged total density of states at the Fermi energy, $n\argu{\varepsilon_F}$. 

Our RDLM program employs the local density approximation (LDA) of DFT within the atomic sphere approximation (ASA). In the language of the KKR method, the ASA together with the use of orientationally averaged densities implies that in any step of the self-consistency procedure, the orientations $\set{\vect e}$ of the local moments are accounted for only by the similarity transformation of the single-site $t$-matrices,
\begin{align}
\underline t_{i\alpha}\argu{\vect e_{i\alpha}}=\underline R\argu{\vect e_{i\alpha}} \underline t_{i\alpha}\argu{\vect e_z} \underline R\argu{\vect e_{i\alpha}}^\dagger,
\end{align}
where $\underline t_{i\alpha}\argu{\vect e_z}$ is the $t$-matrix with exchange field along the $z$ axis, and $\underline R\argu{\vect e_{i\alpha}}$ is the representation of the $SO\!\left(3\right)$ rotation that transforms $\vect e_z$ into $\vect e_{i\alpha}$.

In our calculations an angular momentum cutoff of $\ell_\text{max}=2$ was used for KKR, while the orientational probability was expanded up to $\ell=8$ (cf.\ Eq.~\eqref{eq:h_expansion}) giving adequate convergence even at low temperatures. The discretization of spin directions on the unit sphere was done by using a Lebedev--Laikov grid\cite{lebedev-1999} consisting of 350 points, which was sufficient even in case of peaklike statistical distributions at low temperatures.
Numerical energy integrals were computed along a semicircular contour in the upper complex semiplane using 12 to 16 points depending on the system under consideration.  

\begin{figure*}[ht]
\subfigure{\includegraphics[width=0.45\linewidth]{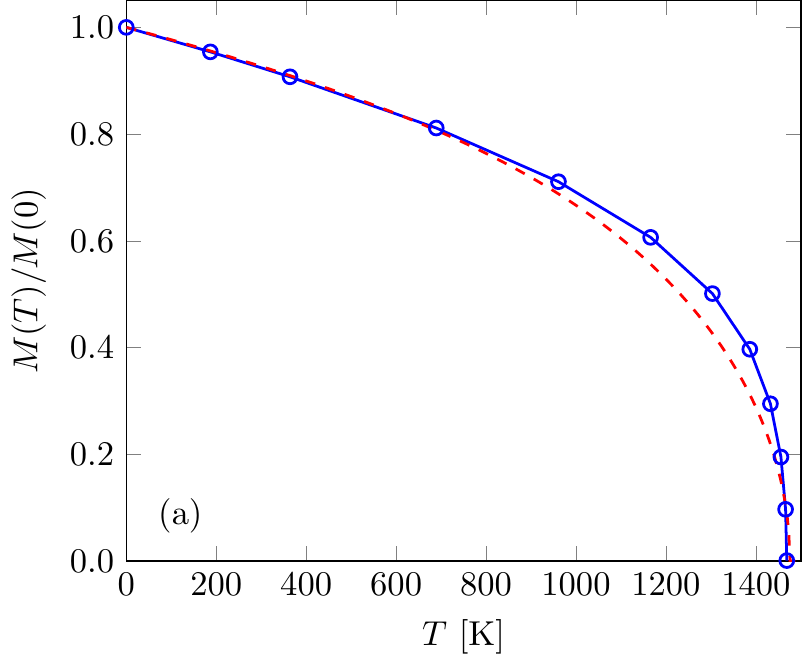}\label{fig:fe_M-T}}
\hfill
\subfigure{\includegraphics[width=0.465\linewidth]{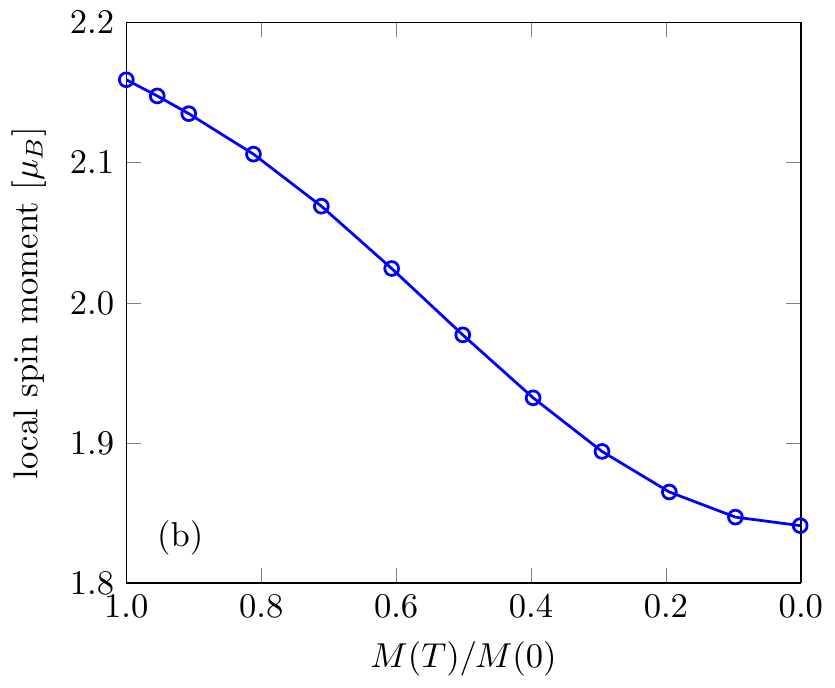}\label{fig:fe_MS-T}}
\caption{(Color online) Temperature dependence of the (a) magnetization (b) local spin moment of bcc Fe obtained from self-consistent RDLM calculations with a lattice constant of $2.79\,\text{\AA}$. The dashed line in (a) shows the mean field solution to a classical Heisenberg model for comparison.}
\end{figure*}

\section{Results}
\label{results}

\subsection{Fe bulk}

We first performed self-consistent RDLM  calculations for bulk bcc Fe with the lattice constant of $2.79\,\text{\AA}$ which is close to the equilibrium value that can be obtained by LDA.\cite{huhne-1998} The dependence of the reduced magnetization, $M(T)/M(0)$ (cf.\ Eq.~\eqref{eq:total_mag}) on the temperature is shown in Fig.~\ref{fig:fe_M-T}. The obtained Curie temperature, $T_\text{C} \simeq 1450 $~K, agrees with earlier DLM results,\cite{buruzs-2008} but it is obviously too high as compared to the experimental value, $T_\text{C} \simeq 1040$~K.  This deficiency can be attributed to the mean field approximation involved in RDLM, and improvements on this approximation such as the use of Onsager cavity fields could provide a more realistic temperature range.\cite{staunton-1992} Nevertheless, since important spin-fluctuations are taken into account in the theory, we emphasize that physical quantities from RDLM calculations should be considered as a function of the magnetization rather than the temperature.

As emphasized in Section~\ref{theory} our present theory allows to calculate the local moments against 
the temperature or average magnetization.  The corresponding results for bcc Fe are shown in Fig.~\ref{fig:fe_MS-T}.
The Fe spin moment of $m_\text{Fe}=2.16 \, \mu_B$ in the ferromagnetic state is in good agreement with other calculations and with experiment, see e.g.\ in Ref.~\onlinecite{huhne-1998}. By increasing the temperature (decreasing the magnetization) $m_\text{Fe}$ monotonously decreases and reaches a value 1.84~$\mu_B$
in the paramagnetic state. This clearly demonstrates that even a system widely regarded as a ``good moment'' one might be subject to considerable longitudinal spin fluctuations at finite temperatures.

The dashed line in Fig.~\ref{fig:fe_M-T} shows a fit of $M(T)/M(0)$ to a classical Heisenberg model in the mean field approximation producing the same Curie temperature. 
Apparently, for higher temperatures the spin model results in significantly lower magnetizations than the RDLM calculations. 
This is in particular surprising since, as discussed above in context to Fig.~\ref{fig:fe_MS-T}, the local moment is even softening with increasing temperatures 
as calculated from the RDLM scheme, while it is \emph{a priori} set to constant within the spin model. 
Still, as our \emph{ab initio} theory does not rely on a spin model, there is no contradiction.
At best, one could try to map the RDLM results by using temperature-dependent spin model parameters. 
Such an attempt has been done by B\"ottcher \emph{et al.}\cite{bottcher-2012} showing an increase of
the dominating nearest-neighbor effective interactions with increasing temperatures, in agreement with our results for the temperature dependence of the magnetization.

\subsection{Magnetic anisotropy of FePt}
The large magnetocrystalline anisotropy of the L1$_0$ FePt alloy and its dependence on temperature and chemical composition has gained large experimental\cite{okamoto-2002,thiele-2002} and theoretical\cite{staunton-2004,staunton-2004b,mryasov-2005a,aas-2013} interest. 
We performed temperature-dependent RDLM calculations by using a lattice parameter $a=2.73\,\text{\AA}$ and a $c/a$ ratio of 0.964.  Similar to previous works\cite{staunton-2004b,aas-2013}  long-range chemical disorder was modelled as intermixing between Fe- and Pt-rich layers. Thus the stacking along the crystallographic $c$ axis consists of alternating nominally Fe layers containing $\eta$ part Fe and $1-\eta$ part Pt, and nominally Pt layers containing $\eta$ part Pt and $1-\eta$ part Fe. The ordered state is described by $\eta=1$, while the completely disordered state corresponds to $\eta=0.5$. The long-range chemical order parameter $S$ can be defined as a linear map between these two extrema as
\begin{align}
S=2\eta-1,
\end{align}
$S=1$ describing the chemically ordered state and $S=0$ meaning complete disorder. Apart from the ordered case we investigated four levels of disorder with $S=0.82$, $S=0.72$, $S=0.62$, and $S=0.52$ to match specific samples in the measurements of Okamoto \emph{et al.}\cite{okamoto-2002} For every case the complete temperature range up to the Curie point and two orientations for the average magnetization were taken into consideration.
In our convention $z$ is parallel to the $c$ axis of the L1$_0$ structure, and the $x$ axis points towards first nearest neighbors in the planes normal to the $c$ axis.

\begin{figure}
\includegraphics[width=\linewidth]{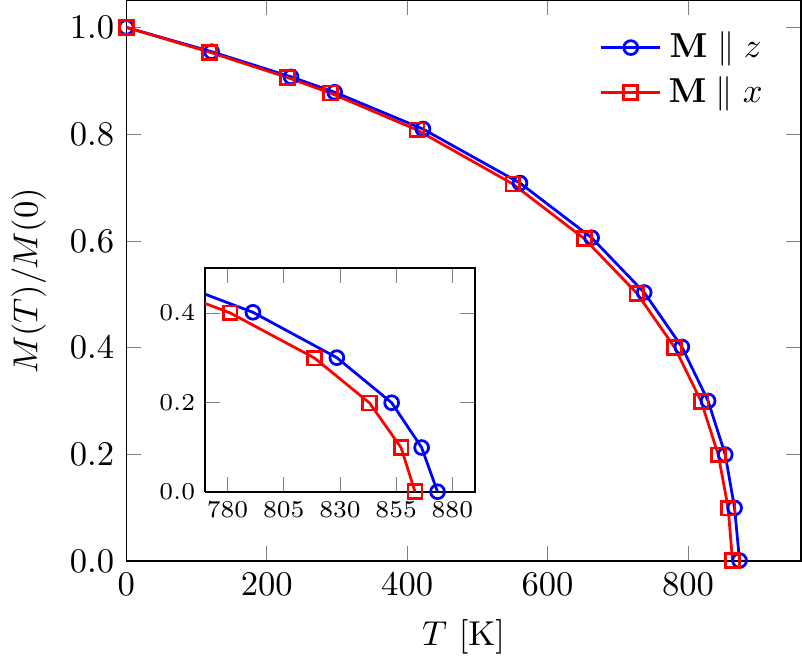}
\caption{\label{fig:fept_M-T}(Color online) Calculated magnetization curves (see Eq.~\eqref{eq:total_mag}) of chemically ordered FePt for magnetization directions along the $c$ axis ($z$) and normal to the $c$ axis ($x$).}
\end{figure}

For both orientations of the magnetization the reduced magnetization per unit cell (i.e., including both Fe and Pt sites) is shown against the temperature in Fig.~\ref{fig:fept_M-T} for the case of perfect chemical order (i.e.,\ no intermixing between the Fe and Pt layers). 
The Curie temperature is found at about 870~K, which is slightly lower than obtained from earlier non-selfconsistent 
calculations\cite{staunton-2004} and it is in fairly good agreement with the experimental value of 750 K.
Noteworthy, $T_\text{C}$ for the $z$ direction of the magnetization is higher than for the $x$ direction.
The shift between the two curves is a clear indication of magnetocrystalline anisotropy. The higher Curie temperature along the $z$ direction indicates that this is the easy axis, in accordance with earlier results. Remarkably, the overall Curie temperature is rather insensitive as the chemical disorder is increased, but the shift between the magnetization curves for $x$ and $z$ becomes gradually smaller, suggesting the decrease of the anisotropy with increasing chemical disorder.
\begin{figure}
\includegraphics[width=\linewidth]{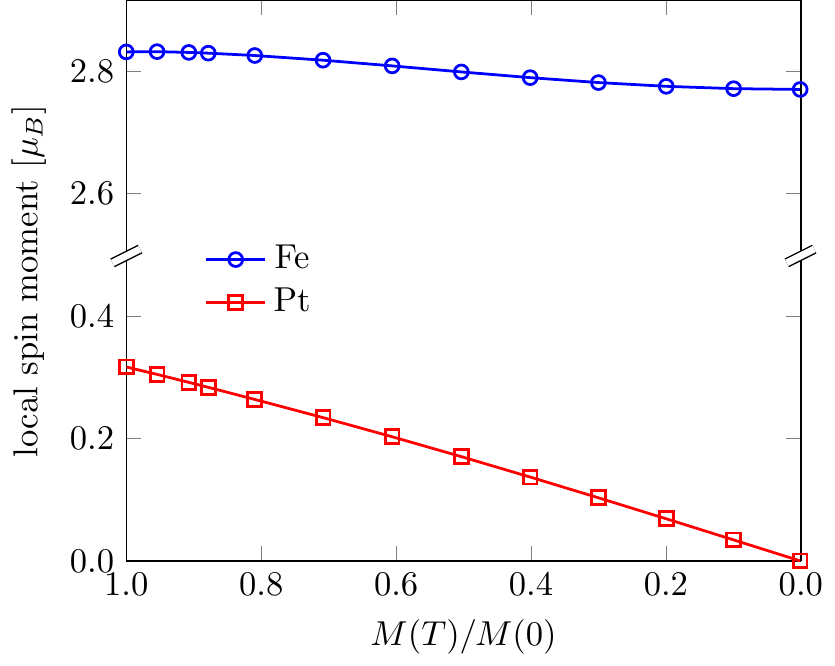}
\caption{\label{fig:fept_MS-T}(Color online) Local spin moments in chemically ordered FePt as a function of the reduced magnetization.}
\end{figure}

The local spin moments at the Fe and the Pt sites, $m_\text{Fe}$ and $m_\text{Pt}$, show completely different behavior against the magnetization as can be seen in Fig.~\ref{fig:fept_MS-T} for the case of chemical order. By increasing the temperature, $m_\text{Fe}$  decreases only by about 2\,\% with respect to its ground state magnitude of $2.83\,\mu_\text{B}$. This implies that the local moment of Fe is more rigid in FePt than in bulk Fe.  
In sharp contrast, the Pt local moment scales with the magnetization of the sample in a very neat linear fashion, reinforcing the simple picture of Pt moments induced by the local Weiss field produced by the Fe moments\cite{mryasov-2005a,mryasov-2005b} (even though no such assumption is involved in the RDLM procedure). Similar behavior is found in chemically disordered systems, there is only a reduction of the zero temperature average Pt local moment from $0.32\,\mu_\text{B}$ (for $S=1$) through $0.28\,\mu_\text{B}$ (for $S=0.72$) to $0.26\,\mu_\text{B}$ (for $S=0.52$).

The temperature-dependent magnetic anisotropy energy $K\argu{T}$ is defined as the difference of the free energies of ferromagnetically ordered systems magnetized along the $z$ and $x$ axes,
\begin{align}
K\argu{T}=F^x\argu{T} -F^z\argu{T},\label{eq:Kscf}
\end{align}
where $F^{x(z)}\argu{T}$ is the free energy of the system magnetized along $x(z)$.
The possibility of treating chemical disorder in our RDLM program in terms of the CPA allows us to improve our theoretical understanding of the magnetic anisotropy in FePt. 
We compute the MAE by employing the magnetic force theorem (MFT).\cite{liechtenstein-1987,jansen-1999} 
Starting with a self-consistent calculation for a system magnetized along the $z$ axis, 
we perform a calculation for the magnetization along the $x$ axis using the same potential and probability distribution. By omitting further self-consistency, 
the total energy difference is approximated by the difference in band energy and, due to the lack of charge conservation, the grand potential should be considered as the
relevant thermodynamic potential,
\begin{align}
\Omega_\text{band}(\eta,T)= & \left\langle E_\text{band}\left(\set{\xi},\set{\vect e}\right)\right\rangle-TS_\text{s}(\eta,T) \notag \\
& - \mu(\eta,T) \left\langle N_\text{v} \left(\set{\xi},\set{\vect e}\right)\right\rangle \, , \label{eq:omega}
\end{align}
with the chemical potential $\mu(\eta,T)$ and the statistical average of the number of valence electrons 
$\left\langle N_\text{v} \left(\set{\xi},\set{\vect e}\right)\right\rangle$. Note that the temperature-dependent part of the electronic free energy, $\Delta F_\textrm{el}(\eta,T)$,
has minor contribution to the MAE, therefore we neglected it from the present calculations. 
As we assume the same probability distribution for spin disorder along the $x$ and $z$ axes, only the first and third terms of the \emph{rhs} of Eq.~\eqref{eq:omega} contribute to the MAE. 
To evaluate the DOS accurately, we used up to 5000 $k$ points in the irreducible wedge of the Brillouin zone near the Fermi energy. 
It is important to note that this approach to calculating the anisotropy clearly fails near $T_\text{C}$, where at a given temperature the probability function 
(therefore, the size of the equilibrium magnetization)
is substantially different for orientations along the $x$ and $z$ direction. The inset of Fig.~\ref{fig:fept_M-T} suggests that 
when the reduced magnetization of the system magnetized along $z$ is below 0.1, the system magnetized along $x$ is even unstable against the paramagnetic state.
It is worth mentioning that within the MFT the MAE can also be calculated by using the magnetic torque.\cite{staunton-2006,buruzs-2007}

Earlier theoretical results showed that the zero-temperature MAE rapidly decreases with chemical disorder, and in the limit of maximal intermixing between Fe and Pt sites the anisotropy almost vanishes as the L1$_0$ structure becomes body-centered tetragonal.\cite{staunton-2004b,aas-2013} Finite-temperature investigations found power-law dependence of the MAE on the magnetization with exponent $2$-$2.1$\cite{staunton-2004,mryasov-2005a} over a wide temperature range, in agreement with experimental observations.\cite{okamoto-2002,thiele-2002} By using self-consistent potentials at finite temperatures and taking into account chemical disorder we may elaborate on the earlier findings of Staunton \emph{et al.}\cite{staunton-2004}

\begin{figure}[ht]
\includegraphics[width=\linewidth]{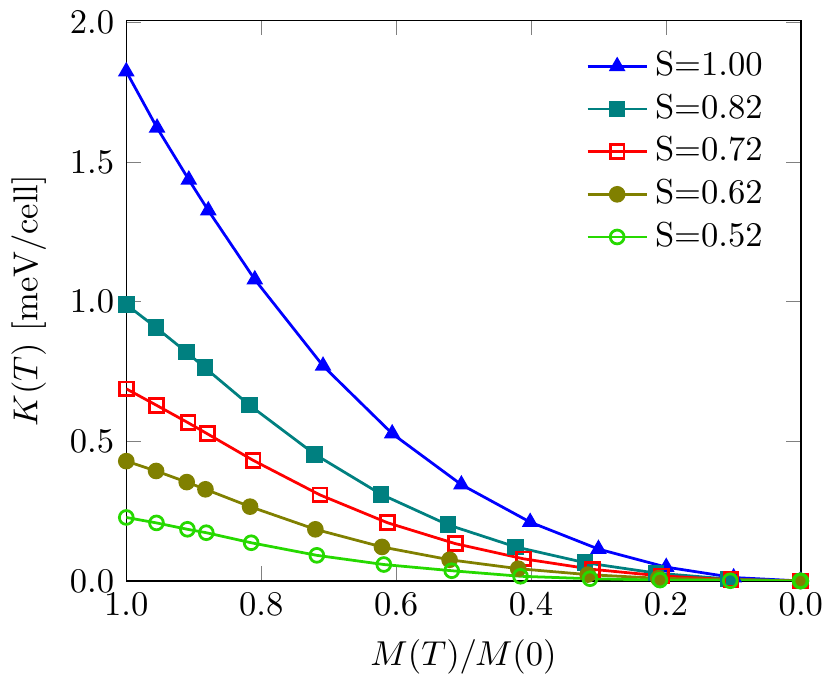}
\caption{\label{fig:fept_MAE-lin}(Color online) Magnetocrystalline anisotropy (cf.\ Eq.~\eqref{eq:Kscf}) versus magnetization in FePt with chemical order parameter $S=1.00$ (chemically ordered state, filled blue triangles), $S=0.82$ (filled teal squares), $S=0.72$ (empty red squares), $S=0.62$ (filled olive circles), and $S=0.52$ (empty green circles).}
\end{figure}

Our results for the MAE per unit cell versus reduced magnetization is shown in Fig.~\ref{fig:fept_MAE-lin} for the five selected values of chemical disorder (positive values indicate that the $z$ axis is favored). The $T=0$~K limit shows the expected rapid decay of the MAE with increasing chemical disorder, as it is reduced from $1.83\,\text{meV}$ ($S=1$) through $0.69\,\text{meV}$ ($S=0.72$) to $0.23\,\text{meV}$ ($S=0.52$).   
\begin{figure}[h]
\includegraphics[width=\linewidth]{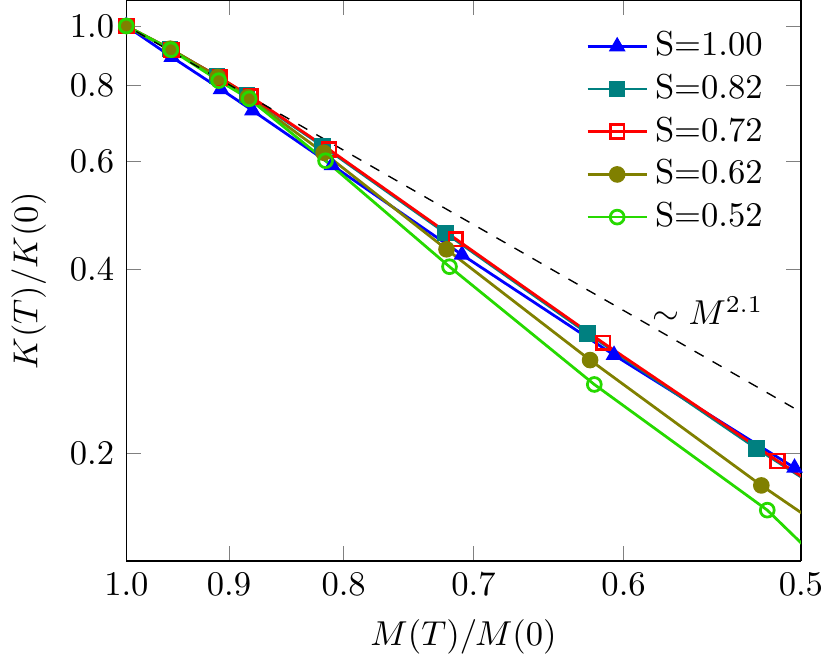}
\caption{\label{fig:fept_MAE-log}(Color online) Magnetization dependence of the reduced anisotropy (cf.\ Eq.~\eqref{eq:Kscf}) in FePt for the five different levels of chemical disorder, each showing power-law behavior. The dashed line indicates the function $M^{2.1}$ for comparison with earlier results in the literature.}
\end{figure}

As clear from Fig.~\ref{fig:fept_MAE-lin}, for each level of chemical disorder the MAE decreases monotonously with increasing temperature (decreasing magnetization) and vanishes at the Curie temperature corresponding to $M\argu{T_\text{C}}=0$. 
For the evaluation of scaling behavior and comparison with the experimental results of Okamoto \emph{et al.} (Ref.~\onlinecite{okamoto-2002} and especially Fig.~9 therein) the reduced MAE curves are shown on a log-log scale in Fig.~\ref{fig:fept_MAE-log}. For ease of comparison with Fig.~9 of Ref.~\onlinecite{okamoto-2002} the shape of the
symbols in that figure are matched in our own for similar values of chemical disorder. All five curves indeed show power-law behavior, and for low temperatures (large magnetizations) they seemingly cluster around $K\sim M^{2.1}$ (dashed line in Fig.~\ref{fig:fept_MAE-log}), as was found by Okamoto \emph{et al}. However, with increasing temperature and chemical disorder the curves gradually drift below this function, which can actually be glimpsed in the data shown in Ref.~\onlinecite{okamoto-2002} as well. We note that the exponent provided by Okamoto \emph{et al}.\ describes low-temperature behavior, and there is little experimental reason to expect uniform power-law behavior up to the Curie point.

It should also be noted that our calculations refer to the total anisotropy energy corresponding to the difference between the magnetization directions along $x$ and $z$, 
whereas Fig.~9 of Ref.~\onlinecite{okamoto-2002} refers to the second order uniaxial anisotropy alone. However, as we checked for the cases of $S=1.00$ and $0.52$, our calculations confirm
negligible ($\lesssim$ 2 \%) higher-order contributions to the MAE. The relatively large $K_2$ found in Ref.~\onlinecite{okamoto-2002} 
might then be attributed to the imperfect film geometry in the experiment (e.g.\ to strain), and our calculated MAE should indeed be compared to the $K_1$ reported there.

\subsection{Metamagnetic transition in FeRh}
The metamagnetic phase transition of FeRh from a high-temperature ferromagnetic (FM) phase to low-temperature antiferromagnetic (AFM) has raised much research. A multitude of theoretical approaches were utilized to gain insight on the nature of the metamagnetic transition, covering first principles total energy calculations,\cite{moruzzi-1992,gruner-2003} \emph{ab initio} spin-fluctuation theory,\citep{sandratskii-1998,kubler-2000} time-dependent excitations,\cite{ju-2004,sandratskii-2012} or effective spin models.\cite{gruner-2003,gu-2005,mryasov-2005b,sandratskii-2011} The DLM theory was shown to accurately predict the metamagnetism of CoMnSi-based alloys.\cite{staunton-2013}
A RDLM scheme, in which the full charge and magnetization self-consistency was approximately accounted for using a comparison between paramagnetic DLM and $T=0\,\text{K}$ FM states, has already been used to determine the free energy of FeRh as a function of different magnetization components, concentrations, external field and temperature, from which the metamagnetic transition temperature and the isothermal entropy change were obtained.\cite{staunton-2014}  

\begin{figure}
\includegraphics[width=\linewidth]{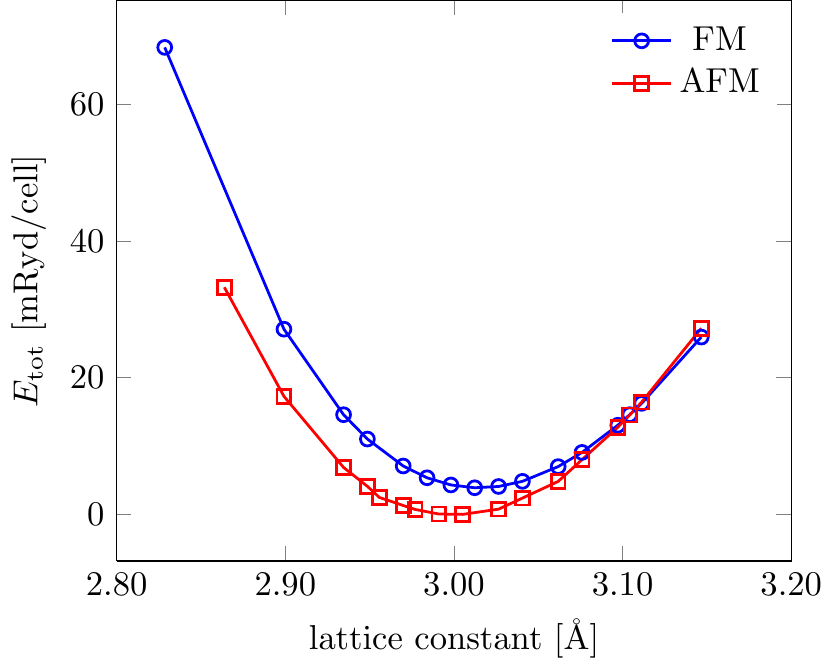}
\caption{\label{fig:ferh_Etot-a}(Color online) Calculated LDA total energies of FeRh as a function of the cubic lattice constant, in the FM (blue circles) and the AFM (red squares) states. The energy scale is normalized to the bottom of the AFM curve.}
\end{figure}

First we performed zero temperature total energy calculations for various values of the lattice constant both in the FM and the bipartite AFM states. As can be see in Fig.~\ref{fig:ferh_Etot-a}, for lattice constants ($a$) less than $3.11\,\text{\AA}$ the AFM state is more stable than the FM one, while for larger lattice constants the FM state becomes more stable. The global energy minimum is found in the AFM state at the equilibrium lattice constant of $a_\text{AFM}=3.00\,\text{\AA}$
and the FM state has an energy minimum at a lattice constant of  $a_\text{FM}=3.02\,\text{\AA}$. To check these results we repeated 
the total energy calculations by using VASP\cite{kresseVASP1,kresseVASP2} and found the AFM and FM energy minima at $2.99\,\text{\AA}$ 
and at $3.01\,\text{\AA}$ respectively, in excellent agreement with the SKKR calculations. 
The corresponding volume increase, $V_\text{FM}/V_\text{AFM}-1=0.02$, is also in fair agreement with earlier theoretical results.\citep{moruzzi-1992,gruner-2003} 
Similarly, the calculated spin moments, $m_\text{Fe}=3.11 \, \mu_B$ and
$m_\text{Rh}=0 \, \mu_B$ in the AFM state, whilst  $m_\text{Fe}=3.22 \, \mu_B$ and
$m_\text{Rh}=1.03 \, \mu_B$ in the FM state, are consistent with values found experimentally\cite{shirane-1963a,shirane-1964} and theoretically.\cite{moruzzi-1992,gruner-2003,sandratskii-2011}
 
\begin{figure}
\includegraphics[width=\linewidth]{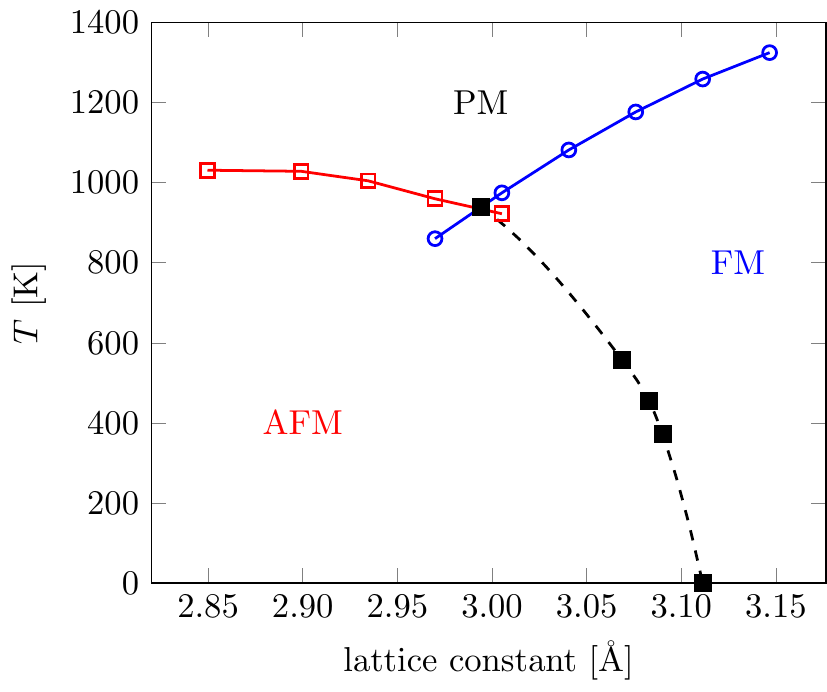}
\caption{\label{fig:ferh_phasediag}(Color online) Approximate phase diagram of FeRh from RDLM as a function of lattice constant and temperature. Blue circles and red squares mark calculated Curie and N\'eel temperatures, respectively, while black filled squares denote metamagnetic transition points as obtained from free energy calculations. The blue and red solid lines and the black dashed lines represent the corresponding phase boundary lines.}
\end{figure}

Next we used the RDLM code to determine the Curie and Néel temperatures respectively for FM and AFM configurations, for a range of lattice constants around the ground state equilibrium values. Note that finding the paramagnetic transition temperature doesn't need a scan over the whole temperature range, since it is sufficient just to set a tiny value of the average magnetization (we usually choose around $M\argu{T}/M\argu{0}=0.01$) and to determine the probability distribution, concomitantly, the temperature that produces the chosen value of the magnetization.  Fig.~\ref{fig:ferh_phasediag} clearly demonstrates  that $T_\text{C}$ and $T_\text{N}$ strongly depend on the lattice constant. At low lattice constants $T_\text{N}$ is larger than $T_\text{C}$, however, with increasing lattice constant $T_\text{N}$ decreases while $T_\text{C}$ increases. This is in accordance with our results for the lattice constant dependence of the total energy (see Fig.~\ref{fig:ferh_Etot-a}) and also with earlier findings that with increasing volume the AFM character of FeRh becomes weaker due to the weakening of the AFM intersublattice Fe-Fe interactions.\cite{sandratskii-2011} Our work on partially ordered and non-stoichiometric FeRh shows that Fe `defects' on the Rh sublattice dramatically enhance the FM interactions.\cite{staunton-2014}

By mapping the dependence of the Néel and Curie temperatures on the lattice parameter, we can gain insight into the high-temperature phase transition describing the system. At low volumes, when $T_\text{C} < T_\text{N}$, there exists a temperature range, $T_\text{C}<T<T_\text{N}$, where the PM phase is stable against the FM phase, however, the AFM state is still ordered, i.e., the PM phase is unstable against the AFM phase. Therefore, in case of  $T_\text{C}<T_\text{N}$ the high-temperature phase transition is AFM-PM.  Conversely, for higher volumes, when $T_\text{N} < T_\text{C}$ ,the high-temperature transition is FM-PM. By scanning the Curie and Néel temperatures as a function of lattice constant we can chart the high-temperature FM-PM and AFM-PM lines of the phase diagram of FeRh, see Fig.~\ref{fig:ferh_phasediag}.

The Néel and Curie temperature curves cross over at $a=2.99\,\text{\AA}$, forming a triple point at $T_\text{N}=T_\text{C}=940\,\text{K}$ (topmost filled black square in Fig.~\ref{fig:ferh_phasediag}). Thus, for $a \ge 2.99\,\text{\AA}$, as annealed from the PM phase the system first orders in the FM phase.
However, the $T=0$ limit (Fig.~\ref{fig:ferh_Etot-a}) indicates that in the ground state the transition from AFM to FM occurs at $a=3.11\,\text{\AA}$ 
(bottommost filled black square in Fig.~\ref{fig:ferh_phasediag}). Consequently, between these two lattice constants 
there has to be an additional transition from the FM to the AFM phase. 
This is the manifestation of the metamagnetic phase transition of FeRh provided by our RDLM theory. 
We note that the situation is similar to the spin-reorientation transitions in ferromagnetic thin film systems: 
when the ground-state magnetization is oriented normal to plane, while the Curie temperature related to the 
in-plane magnetization is higher than for the out-of-plane magnetization, 
a temperature-induced reorientation transition occurs between these two orientations.\cite{udvardi-2001}

To find the metamagnetic transition line of the phase diagram, we computed the RDLM free energy curves for some values of the lattice constant. As mentioned above, in the AFM phase the Weiss field at the Rh sites vanishes and local Rh moments only form in the FM phase. These moments were treated as induced in our calculations, and as such enslaved to the robust Fe moments (the magnitude of which changed less than 2\,\% as a function of temperature for every case). The spin-disorder entropy term Eq.~\eqref{eq:Sc} entering the free energy correspondingly only contains contributions from the Fe sites, however the cost of Rh moment formation is included in the (average) total energy in a self-consistent manner.
 
\begin{figure}
\includegraphics[width=.95\linewidth]{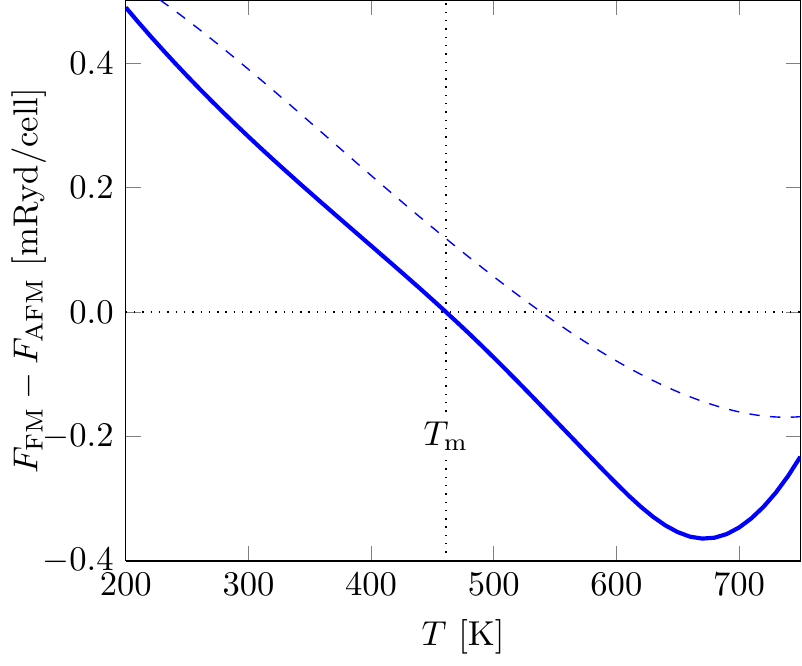}
\caption{\label{fig:ferh_crossover}(Color online) 
Calculated free energy difference (see Eq.~\eqref{eq:F}) between the FM and AFM states 
at $a=3.08\,\text{\AA}$ displaying a crossover at $T_\text{m}=461\,\text{K}$. The dashed curve corresponds to the case when the electronic free energy contribution,
$\Delta F_\text{el}\argu{T}$, is neglected.} 
\end{figure}

Reassuringly, for lattice constants $a<2.99\,\text{\AA}$ and $a>3.11\,\text{\AA}$ we didn't find a crossover between the FM and AFM free energy curves, indicating that one of these phases remains stable in the entire temperature range up to the PM transition temperature.  For the case of $a=3.08\,\text{\AA}$ we plotted the difference of the free energies, $F_\text{FM}\argu{T}-F_\text{AFM}\argu{T}$, in Fig.~\ref{fig:ferh_crossover} for a broad temperature range. Note that this plot is in fact the difference of cubic spline fits to the calculated free energies, since the AFM and FM RDLM calculations were performed for different temperature values. This plot indeed reveals a crossover through zero at temperature $T_\text{m}=461\,\text{K}$, which can be interpreted as the metamagnetic transition temperature at this fixed lattice constant. Noteworthy, for high temperatures the free energy difference turns back towards zero. This happens since both the FM and AFM states approach smoothly the PM state, implying
\begin{align}
\left. F_\text{FM}\right\rvert_{T\to T_\text{C}}=\left. F_\text{AFM}\right\rvert_{T\to T_\text{N}}=E_\text{PM}-T S_\text{PM}+\Delta F_\text{el}\argu{T} \, , \label{eq:f_pm} 
\end{align}
$E_\text{PM}$, $S_\text{PM}$  and $\Delta F_\text{el}\argu{T}$ being the average total energy, the spin-entropy and the excess electronic free energy (cf.\ Eq~\eqref{eq:F_el}) in the paramagnetic phase, respectively. 

We also examined the role of the electronic free energy, $\Delta F_\text{el}\argu{T}$ (Eq.~\eqref{eq:F_el}), in the metamagnetic phase transition. 
The dashed curve in Fig.~\ref{fig:ferh_crossover} corresponds to the case when $\Delta F_\text{el}\argu{T}$ was neglected. 
As can be seen, switching off $\Delta F_\text{el}\argu{T}$ increases the free energy of the FM phase with respect to the AFM phase in the whole temperature range.
This can clearly be attributed to the fact that the DOS at the Fermi energy is significantly larger in the FM phase than in the AFM phase. Only above 650~K does
$n_\textrm{AFM}(\varepsilon_F)$ start to increase rapidly, giving rise to a sharp decrease of the magnitude of the free energy difference between the two phases.
As a consequence, due to $\Delta F_\text{el}\argu{T}$ $T_\textrm{m}$ decreases from 539~K (zero of the dashed curve) to 461~K. 

We completed calculations to determine the metamagnetic transition temperature for three selected lattice constants. The corresponding $T_\text{m}\argu{a}$ data are marked in Fig.~\ref{fig:ferh_phasediag} as black filled squares. Nevertheless, including the $T=0$ point at $a=3.11 \, \text{\AA}$ and the triple point, $T=940 \, \text{K}$
at $a=2.99 \,  \text{\AA}$, a smooth metamagnetic phaseline could be drawn (black dashes) making the RDLM phase diagram complete. 

While the existence of metamagnetic phase transitions as well as the trend of magnetic ordering against the volume   
are correctly captured by our RDLM theory, the quantitative agreement is rather poor with respect to experiments
reporting  $T_\text{m}=338 \,\text{K}$\cite{shirane-1963b} and $T_\text{C}=678 \,\text{K}$\cite{shirane-1964} at zero pressure  ($a \simeq 2.99 \text{\AA}$).  
As already discussed before, one reason for this disagreement can be understood  due to the mean field approximation overestimating the transition temperatures. A similarly high Curie temperature, $T_\text{C}=885 \,\text{K}$,  at $a=2.98 \,  \text{\AA}$ was obtained also from \emph{ab initio} spin-fluctuation theory.\cite{sandratskii-1998,kubler-2000} Another shortcoming of the phase diagram in Fig.~\ref{fig:ferh_phasediag} is that the region of the equilibrium lattice constants, $3.00\,  \text{\AA} \le a \le 3.02\,  \text{\AA}$ is very close to the triple point at $a=2.99\,  \text{\AA}$ resulting in too high metamagnetic transition temperatures of about 800-900~K. In terms of spin-fluctuation theory~\cite{sandratskii-1998,kubler-2000} a more realistic transition temperature of $T_\text{m}=435\,  \text{K}$ was obtained. This could indicate that the spin-disorder entropy is underestimated by our RDLM approach in the FM state of FeRh, most probably because of neglecting the transversal degrees of freedom of the rather large induced spin-moments of Rh. It should be mentioned that the present RDLM results could also be improved by using exchange-correlation functionals beyond LSDA like the generalized gradient approximation (GGA).\cite{gu-2005} The total energy differences between the FM and AFM states could also be inaccurate due to the ASA, and so a full potential treatment should improve our results. Furthermore, our recent finding of the high compositional sensitivity of the metamagnetic transition highlights a possible source for the disparity between experiments and theory.\cite{staunton-2014}

\section{Conclusions}
\label{conclusions}
We have presented a relativistic disordered local moment scheme capable of describing finite-temperature spin disorder self-consistently. Based on the Feynman--Peierls--Bogoliubov inequality a variationally best mean field approximation provides the orientational distribution of spins, which is iterated simultaneously with the potentials during the self-consistent loop. The KKR-CPA method provides a convenient and natural framework for the theory. Using a self-consistent procedure at finite temperatures gives us a powerful tool by including longitudinal spin fluctuations and the effect of induced moments on the electronic structure. Relativistic effects are included as the scattering problem is described by the Dirac equation.

Test calculations for bulk bcc Fe showed that even in a ``good moment'' system the magnitude of the local moment can vary significantly with spin disorder. Because of the mean field approximation underlying the RDLM scheme the Curie temperature was largely overestimated by the calculations. Therefore, the obtained thermodynamic quantities should be considered as a function of the average magnetization rather than of the temperature.

The temperature-dependent calculations revealed that in the FePt alloys the spin moment of Fe is stable within about 2\,\%, while Pt displays an induced moment indicated by its linear relationship with the overall magnetization of the system. In agreement with earlier results we established that the magnetocrystalline anisotropy of FePt is drastically reduced by increasing long-range chemical disorder. The magnetization dependence of the MAE was found power-like for any degree of chemical disorder, with an exponent of about 2.1 for weak disorder and of somewhat larger value for more more disordered samples, reproducing the tendencies reported by experiment.

We set up an \emph{ab initio} phase diagram for bulk FeRh as a function of the lattice constant and the temperature.
The tendency of AFM vs.\ FM order against volume was found correctly both for the ground state and for the paramagnetic transitions. From simple thermodynamic arguments we concluded that there is a range of volume where a temperature-induced  AFM-FM phase transition should occur.    
The temperature of the metamagnetic transition were numerically determined  from the crossovers in the free energy of the FM and AFM states. 

Beside of the qualitative success of our RDLM theory, we noticed quantitative disagreements, in particular, concerning the estimated transition temperatures.  These can be partly attributed to the mean field treatment of the spin fluctuations. In addition, in our present implementation we neglected the effect of orientational spin-fluctuations on the Kohn--Sham potentials and fields, as we calculated them from the statistically averaged Green's function (cf.\ Eqs.~\eqref{eq:avgrho}--\eqref{eq:avgmag}).
An obvious point for development of our method is, therefore, to consider orientation-dependent densities and potentials at the sites with fluctuating spin-moments. This improvement could affect the single-site probability distributions and consequently the obtained self-consistent states, as well as the corresponding temperatures. 
This can remarkably be reflected in the results for subtly balanced systems such as FeRh with FM-AFM instability.

\begin{acknowledgments} 
Illuminating discussions with L\'aszl\'o Udvardi are kindly acknowledged.
Financial support was provided by the Hungarian National Research Foundation (under contracts OTKA 77771 and 84078), and in part by the European Union under FP7 Contract No.\ NMP3-SL-2012-281043 FEMTOSPIN (A.D., E.S., L.B., L.S.), and by the EPSRC (UK) grant EP/J006750/1 (J.B.S).
M.S.D.\ acknowledges support of the HGF-YIG Programme VH-NG-717.
The work of L.S.\ was also supported by the European Union and the State of Hungary, co-financed by the European Social Fund in the framework of T\'AMOP 4.2.4.\ A/2-11-1-2012-0001 `National Excellence Program'.

\end{acknowledgments}

\end{document}